\documentclass[aps,prl,twocolumn,superscriptaddress]{revtex4-2}
\usepackage{graphicx}
\usepackage{siunitx}
\usepackage{easyReview}
\usepackage[colorlinks=true, allcolors=blue]{hyperref}

\begin{document}

%\preprint{}

\title{Comparative structural evolution under pressure of powder and single crystals of the layered antiferromagnet FePS$_3$}


\author{David M. Jarvis}
\email[Corresponding author: ]{jarvis@ill.fr}
%\thanks{}
\affiliation{Institut Laue-Langevin, 71 Avenue des Martyrs, 38000 Grenoble, France}
\affiliation{Cavendish Laboratory, Cambridge University, J.J. Thomson Ave, Cambridge CB3 0HE, UK}

\author{Matthew J. Coak}
\affiliation{Cavendish Laboratory, Cambridge University, J.J. Thomson Ave, Cambridge CB3 0HE, UK}
\affiliation{Department of Physics, University of Warwick, Gibbet Hill Road, Coventry CV4 7AL, UK}

\author{Hayrullo Hamidov}
\affiliation{Cavendish Laboratory, Cambridge University, J.J. Thomson Ave, Cambridge CB3 0HE, UK}
\affiliation{Navoi State Mining Institute, 72 M. Tarobiy Street, Navoi, 210100, Uzbekistan}

\author{Charles R.S. Haines}
\affiliation{Physics, University of East Anglia, Norwich, NR4 7TJ, UK}
\affiliation{Cavendish Laboratory, Cambridge University, J.J. Thomson Ave, Cambridge CB3 0HE, UK}

\author{Giulio I. Lampronti}
\affiliation{Department of Earth Sciences, University of Cambridge, Downing Street, Cambridge CB2 3EQ, UK}

\author{Cheng Liu}
\affiliation{Cavendish Laboratory, Cambridge University, J.J. Thomson Ave, Cambridge CB3 0HE, UK}

\author{Shiyu Deng}
\affiliation{Cavendish Laboratory, Cambridge University, J.J. Thomson Ave, Cambridge CB3 0HE, UK}

\author{Dominik Daisenberger}
\affiliation{Diamond Light Source, Chilton, Didcot OX11 0DE, United Kingdom}

\author{David R. Allan}
\affiliation{Diamond Light Source, Chilton, Didcot OX11 0DE, United Kingdom}

\author{Mark  R. Warren}
\affiliation{Diamond Light Source, Chilton, Didcot OX11 0DE, United Kingdom}

\author{Andrew R. Wildes}
\email[Corresponding author: ]{wildes@ill.fr}
\affiliation{Institut Laue-Langevin, 71 Avenue des Martyrs, 38000 Grenoble, France}

\author{Siddharth S. Saxena}
\email[Corresponding author: ]{sss21@cam.ac.uk}
\affiliation{Cavendish Laboratory, Cambridge University, J.J. Thomson Ave, Cambridge CB3 0HE, UK}
%\affiliation{National University of Science and Technology \textquotedblleft MISiS\textquotedblright , Leninsky Prospekt 4, Moscow 119049, Russia}


\date{\today}

\begin{abstract}
The layered antiferromagnet FePS$_3$ has been shown to undergo a structural transition under pressure linked to an insulator-metal transition, with two incompatible models previously proposed for the highest-pressure structure. We present a study of the high-pressure crystal structures of FePS$_3$ using both single-crystal and powder x-ray diffraction. We show that the highest pressure transition involves a collapse of the inter-planar spacing of this material, along with an increase in symmetry from a monoclinic to a trigonal structure, to the exclusion of other models. The extent of this volume collapse is shown to be sensitive to the presence of a helium pressure medium in the sample environment, indicating that consideration of such experimental factors is important for understanding high-pressure behaviours in this material.
\end{abstract}

%\keywords{}

\maketitle

%\section{Introduction}

The layered antiferromagnetic insulators $M$P$S_3$, where $M$ is a first-row transition metal, provide an ideal space in which to explore the behaviour of quasi-two-dimensional magnetic systems and the correlated electron physics within them.

The compounds share a common monoclinic structure, space group $C2/m$, consisting of near-perfect honeycomb layers of  transition metal ions in the $ab$ planes\cite{ouvrard1985,brec1986}. These planes are weakly coupled to each other by van der Waals forces. Magnetic exchange, partly mediated by the P$_2$S$_6$ clusters, is also much stronger within the planes than between them, giving a good approximation of a two-dimensional magnetic system.

Studied extensively since the 1970s, the magnetic exchanges and anisotropies in these materials may be varied by the selection of the transition metal species\cite{ouvrard1985,brec1986}. The extreme two-dimensional limit has been approached through delamination towards a single monolayer, with FePS$_3$ itself found to retain an Ising-type antiferromagnetic order down to the monolayer limit\cite{lee2016,kuo2016,wang2016}.

Complementary to this, a reduction of the inter-layer spacing may give rise to a more three dimensional structure, providing access to an essential area of interest for both magnetic and transport properties on this crossover. The application of hydrostatic pressure is an excellent, targeted technique for tuning in this manner, acting primarily to reduce the separation of the weakly-coupled transition metal planes.

Previous high-pressure work\cite{haines2018,zheng2019,evarestov2020,coak2021} has demonstrated the occurrence of two structural transitions under pressure in powder samples of FePS$_3$. The first, at a critical pressure of \SI{4}{\giga\pascal}, is characterised as a shear of the weakly-coupled $ab$ planes along the crystallographic $\mathbf{a}$ direction, maintaining the ambient pressure $C2/m$ space group whilst reducing the angle $\beta$ to very nearly \ang{90}\cite{haines2018}.

The second transition, at a pressure of \SI{\sim14}{\giga\pascal}, has been found to involve a significant volume collapse of the unit cell, which is also coincident with an insulator-to-metal transition in the compound\cite{haines2018,wang2018a,zheng2019}. The origin of this volume collapse, however, remains the subject of differing interpretations from powder x-ray diffraction experiments. One model attributes this volume reduction to a collapse of the inter-planar spacing with a simultaneous symmetry increase to a trigonal $P\bar{3}1m$ space group\cite{haines2018}; the other to a shrinking of the Fe$^{2+}$ honeycombs, an effect within the $ab$ planes, maintaining a monoclinic structure\cite{wang2018a}. This impact of this higher-pressure structural transition on the magnetism in the system is also the subject of contradictory findings, with indirect results from x-ray emission spectroscopy measurements\cite{wang2018a} contradicted by neutron diffraction\cite{coak2021}. 

 The crystal structure of this high-pressure phase provides the starting point for theoretical calculations and models of the novel magnetic and electronic properties, and so the correct identification of the structural transitions in this compound is of vital importance for understanding other behaviours in this and related materials.
 
 Experimental reports of the high-pressure crystal structures of FePS$_3$ have thus far been determined from powder samples, whereas reports of high-pressure bulk properties, such as measurements of resistivity, have been performed using single-crystals\cite{haines2018,wang2018a}. Verification of the high-pressure behaviour of the crystal structure in single crystals is therefore a necessary step for consolidating understanding of behaviours in the system.
%TODO Add references.

We report here a study examining the evolution of the crystal structure under pressure of FePS$_3$ by diffraction of synchrotron radiation. We show new data from both single crystal and powder samples in diamond anvil cells with a helium pressure transmitting medium, and compare these new data with  previously published measurements\cite{haines2018}  of powder samples without a pressure medium. The highest-pressure transition is found in all cases to include an increase in symmetry to a trigonal $P\bar{3}1m$ structure and a collapse of the inter-planar spacing in both powder and single crystal studies. The extent of this $c$-axis collapse is found to be sensitive to the presence of a helium pressure-transmitting medium, being greatest in experiments without such medium, showing that careful consideration of the sample environment is necessary for the understanding of the high-pressure behaviour of this material.

\subsection*{Methods}

Single crystals of FePS$_3$ were grown by a vapour transport method as described in detail in reference \cite{lancon2016} using stoichiometric quantities of Fe, P and S powders (all of purity $\geq\SI{99.998}{\percent}$).

Samples of FePS$_3$ were manually delaminated down to a thickness of \SI{\sim 15}{\micro\metre} and cut by scalpel and razor blade to squares of side length \SI{\sim 80}{\micro\metre}.  Single crystal x-ray diffraction data was collected on the I19-2 beamline at the Diamond Light Source. An incident beam of wavelength \SI{0.4859}{\angstrom} ($E=\SI{25}{\kilo\electronvolt}$) was used to collect diffraction patterns using a Dectris Pilatus 300K detector. The platelet samples were mounted such that the $\mathbf{c^{*}}$ axis was approximately parallel to the incident beam. Data was then collected as the crystal was rotated about $\phi$ and $\omega$ in a standard configuration. The data was analysed using Crysalis Pro software\cite{crysalis}. %TODO - Finish rotation description.

Powder samples were prepared by grinding the as-grown crystals under liquid nitrogen in order to mitigate the noted tendency towards strong preferred orientations in these materials, arising from the formation of small platelets from shear of the $ab$ planes. This preparation is identical to that used in previous work\cite{haines2018} in order to allow for accurate comparison of other experimental factors in powder experiments.

High-pressure powder x-ray diffraction measurements were performed on the I15 Extreme Conditions beamline at the Diamond Light Source. X-rays of wavelength \SI{0.4246}{\angstrom} ($E=\SI{29.2}{\kilo\electronvolt}$) were used with a MAR345 area detector. Powder data were initially processed using D\textsc{awn} software\cite{filik2017} and Rietveld and Le Bail refinements were performed using TOPAS\cite{topas} and GSAS-II\cite{gsasii}.

In both experiments, the gasket material used with the diamond anvil cells was rhenium, and the cells were loaded with helium gas as a pressure-transmitting medium, which was not used for the previously published experiment\cite{haines2018}. The pressure inside the sample space was determined by measurement of the fluorescence of ruby spheres inserted alongside the sample in the pressure region\cite{mao1986}. 

Diffraction peaks arising from the diamond anvils were identified by fitting the known diamond unit cell and removed from subsequent treatment. Furthermore for the single-crystal experiment, regions of detector images containing powder-like rings from the rhenium gasket were also excluded. Such features remain visible in the generated diffraction images but were excluded from refinements.

\subsection*{Results}

\begin{figure*}
	\centering
	\includegraphics[width=\linewidth]{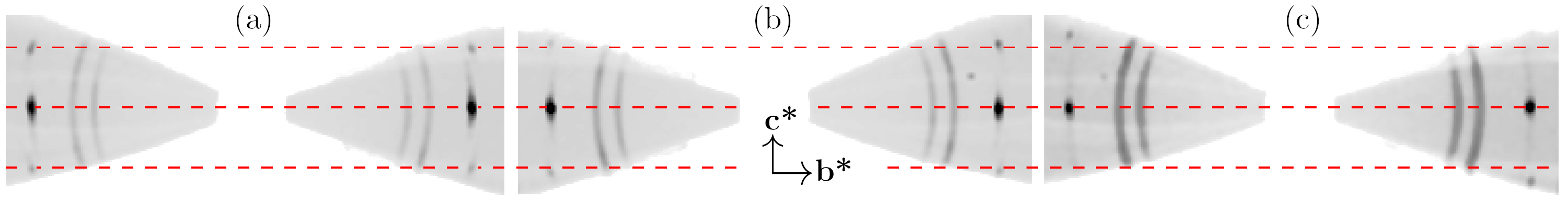}
	\caption{\label{fig:0klComparison}Reciprocal space image of the ($0kl$) scattering plane in single crystal FePS$_3$ in the three distinct structural phases at (a) \SI{1.2}{\giga\pascal}, (b) \SI{6.7}{\giga\pascal}, (c) \SI{19.1}{\giga\pascal}. Visible spots in the lower two pressures are indexed with $k=\pm6$ and in the highest pressure as $k=\pm3$. The dashed lines from bottom to top indicate the lowest pressure position of the $(0k\bar{1})$, $(0k0)$, and $(0k1)$ lines, illustrating the increase in $Q_{(001)}$ with pressure.}
\end{figure*}

%TODO - fill in the correct supplementary figure.
Both the new powder diffraction (using a helium pressure medium) and single crystal data confirm the presence of two structural phase transitions. The first transition from the ambient to the first high-pressure phase, hereinafter referred to as HP-0 and HP-I respectively, occurred at \SI{4}{\giga\pascal}. The transition to the second phase, denoted HP-II, occurred at \SI{14}{\giga\pascal}. Unlike the powder data without a pressure medium\cite{haines2018}, all the transitions were abrupt with no clear evidence of phase coexistence. The powder data and refinements are shown in the supplementary information.

Analysis of both the new powder data and the single crystal data confirmed the conclusions of reference \cite{haines2018}, that the HP-I phase had a space group $C2/m$ with $\beta$ close to \ang{90} and the HP-II phase had a higher symmetry trigonal $P\bar{3}1m$ space group. Unit cell parameters for each phase are given in the supplementary information.

The $(00l)$ peaks were not accessible in the single crystal measurements due to the experimental geometry. To verify the indexing of the single crystal Bragg peaks, and to independently verify the pressure evolution of the lattice parameters, slices of the $(0kl)$ planes were extracted. The $0k0$ and $00l$ directions are orthogonal in all the considered space groups. The $0kl$ Bragg peaks thus give unambiguous measures of the inter-planar spacing, and of a characteristic intra-planar spacing, of the lattice as a function of pressure.
%TODO - Added paragraph, check how it sounds here.

%Two structural transitions are observed in single crystal diffraction at pressures corresponding to those reported in powder studies\cite{haines2018}. The first transition is seen to occur at a critical pressure of \SI{4}{\giga\pascal}, from the ambient phase to a phase hereinafter referred to as HP-I; and a second at a pressure of \SI{14}{\giga\pascal} to a phase denoted HP-II. 
%TODO - This probably still needs some evidence.

%From single-crystal diffraction data, the HP-I phase is fit with a $C2/m$ unit cell with the angle $\beta$ equal to \ang{\sim90}. The higher pressure HP-II phase is fit with a higher symmetry trigonal-hexagonal $P\bar{3}1m$ space group. Refined unit cells corresponding to these two phases are given in the supplementary information.
%TODO - Talk about integrated intensity problems.

Figure \ref{fig:0klComparison} shows slices of the ($0kl$) scattering planes generated from single crystal diffraction data at three pressures corresponding to the determined HP-0, HP-I and HP-II phases.
%These images show the accessible window of reciprocal space with $(0k0)$ the horizontal and $(00l)$ the vertical direction, both being zero at the image centres.
The strong monoclinic $(06l)$ and $(0\bar{6}l)$ peaks for $l=-1,0,1$ which are instead indexed in HP-II as $(0,\pm3,l)$, are visible alongside powder rings from the rhenium gasket. Smearing of the strongest Bragg peaks along the $00l$ direction is attributable to stacking faults and the quasi-two-dimensional nature in this family of materials which have been previously observed\cite{goossens2011,wildes2015,murayama2016}. These slices show a distinct increase in the separation of the peaks along the $(00l)$ direction between the HP-I and HP-II phases, corresponding to a reduction in the distance between the transition metal planes, without a similar change in the $(0k0)$ distances. This is immediate evidence that the volume reduction upon the transition to HP-II is due to changes between the planes, rather than an intra-planar change of the honeycombs  themselves.

To allow for effective quantitative comparison of the crystal structure across the transitions and in different space groups, two key quantities are determined from both powder and single crystal to characterise changes in distances both between the van der Waals planes and between individual Fe$^{2+}$ ions within them.

Inter-planar distance
%in both the $C2/m$ and the $P\bar{3}1m$ structures
may be determined from powder diffraction patterns directly from observation of the $(001)$ diffraction peak. Such analysis however relies on the correct indexation of diffraction peaks. Indexation of powder diffraction patterns of the HP-II phase is found to be possible using a $C2/m$ space group as for the lower pressure phases, but this leads to a misidentification of the important $(001)$ peak and of the overall structure which is clearly incompatible with the structure identified from single-crystal. This is demonstrated fully in the supplementary information, as is the quantitative treatment of reciprocal space images such as those in Figure \ref{fig:0klComparison} for the determination of an equivalent inter-planar distance.

\begin{figure}
	\centering
	\includegraphics[width=\linewidth]{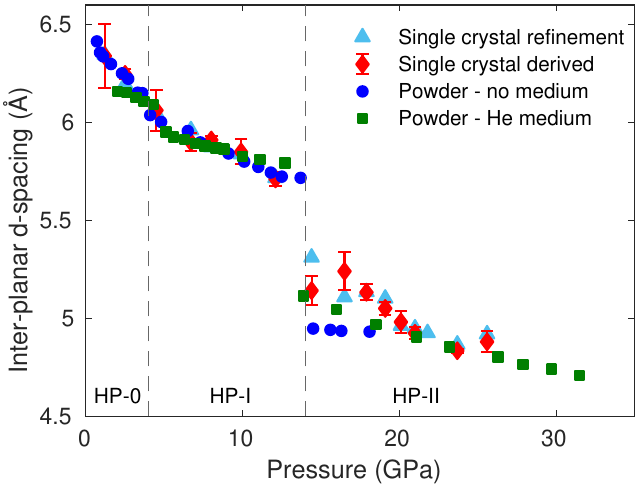}
	\caption{\label{fig:SingleXtalvsPowder001}d-spacing of the (001) diffraction peak, corresponding to the inter-layer separation in the material in (cyan) single crystal refinement; (red) single crystal as determined from $(0kl)$ images; (green) powder sample with a He pressure medium; (blue) powder x-ray diffraction data without pressure medium of Haines et al.\cite{haines2018}}
\end{figure}

%In Figure \ref{fig:0klComparison}, no $(00l)$ peaks are observable due to the orientation of the single crystal sample, as the incident peak is almost normal to $\mathbf{c^*}$. The $(001)$ $d$-spacing is necessarily determined from the overall refinement but is also verified directly from reciprocal space images such as those in Figure \ref{fig:0klComparison} by fitting of the positions in $Q$ of the visible $(0,\pm6,0)$ and $(0,\pm6,\pm1)$ peaks as well as similar pairs related by symmetry and the conversion of this to an equivalent real-space distance.
Figure \ref{fig:SingleXtalvsPowder001} shows this inter-planar spacing in FePS$_3$ as a function of pressure across the three distinct structural phases from: automated refinements of our single crystal data using Crysalis Pro; as determined from analysis of $0kl$ reciprocal space slices; and from Le Bail refinements of powder diffraction data both with and without a helium pressure medium. The transition from HP-0 to HP-I is not clearly observed as this transition involves a shear along the a-axis, with a minimal change in the inter-planar spacing, but it is apparent in a shift in intensities around the $(131)$ and $(20\bar{2})$ peaks as shown in the supplementary information in Figure S4. The transition to HP-II is observed at similar pressures across the different experiments, and in all cases this transition involves a reduction of the inter-planar distance. The extent of this collapse differs between experiments.  The powder measurement without pressure-transmitting medium shows the greatest change of \SI{13.5}{\percent}, from \SIrange{5.718}{4.948}{\angstrom}, whilst the powder measurement with helium medium shows a reduction of \SI{11.7}{\percent}, from \SIrange{5.794}{5.114}{\angstrom}. The single crystal measurement, also with helium medium, shows a reduction between \SI{7.0}{\percent} and \SI{11.3}{\percent}.

\begin{figure}
	\centering
	\includegraphics[width=\linewidth]{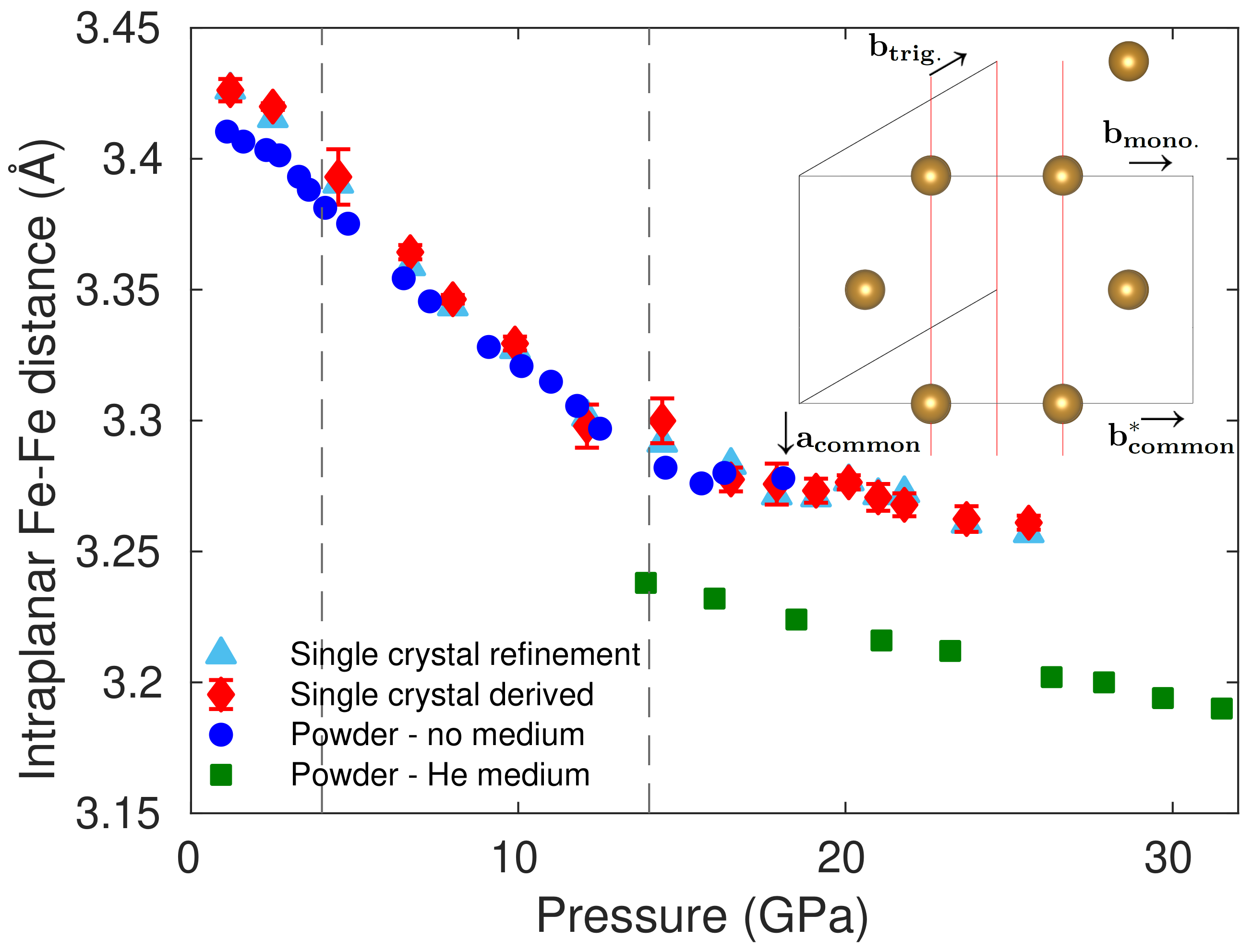}
	\caption{\label{fig:Fe-Fe_distance}Average intra-planar nearest neighbour distance between Fe ions as a function of pressure as determined from (cyan) single crystal refinement; (red) single crystal as determined from $(0kl)$ and equivalent images; (green) powder sample with a He pressure medium; (blue) powder diffraction without pressure medium. (Inset) view along $\mathbf{c^*}$ of the monoclinic and trigonal unit cells, with the spacing of the $(060)$ monoclinic or $(030)$ trigonal lattice planes in red.}
\end{figure}

The structural evolution within the $ab$ planes may be characterised in terms of a nearest-neighbour distance between Fe ions in the honeycombs which is shown in Figure \ref{fig:Fe-Fe_distance}. For single crystal data this distance is derived from the position of the $(060)$ and $(0\bar{6}0)$ peaks as indexed in the monoclinic unit cell, or the $(030)$ and $(0\bar{3}0)$ in the trigonal HP-II cell. The Fe atoms form an almost perfect honeycomb lattice in the HP-0 phase\cite{ouvrard1985} and as shown in the inset of Figure \ref{fig:Fe-Fe_distance},  in the limit of a perfect honeycomb structure the spacing of these $(0k0)$ planes is equal to half the side length of the Fe ion hexagons. Due to correlations observed in refinements for the powder measurement with helium medium (see supplementary information), only values for the HP-II phase are shown.
%The other side lengths of these hexagons, being different in the monoclinic phases where they are slightly distorted, are determined from other symmetry related pairs of diffraction peaks, and an average is shown.

Figure \ref{fig:Fe-Fe_distance} shows that the characteristic intra-planar spacing does not display a significant discontinuity across either of the transitions in any of the experiments performed. A discontinuity is observed in the gradient of this value with pressure across the HP-II transition, with the in-plane distance being more resistant to compression with pressure in the HP-II phase. The absence of a discontinuous change, combined with the observation of the collapse shown in Figure \ref{fig:SingleXtalvsPowder001}, shows unambiguously that the volume collapse is due to a discontinuous change in the inter-layer spacing. The overall reduction of the intra-planar spacing in the HP-II phase is significantly greater in the powder diffraction data measured with a helium pressure medium than in the three other experiments.

%Smaller differences between experiments are observed in the inter-planar distances in the HP-II phase. In this case samples measured with a helium pressure medium remain consistently above those measured without, up to the maximum pressure of the no-medium data. The values are not in agreement taking into account the determined uncertainties from the different experiments. The impact of pressure medium on the transition is then a reduction of the inter-planar collapse, whilst not changing the observed symmetry increase to a trigonal-hexagonal structure.

%  These distances are determined from the positions of the ($0\bar{6}0$) and ($060$) peaks in the $C2/m$ indexing, and equivalently the ($0,\pm3,0$) in the $P\bar{3}1m$.

%TODO - Is this paragraph definitely cut?
%Coexistence of the phases across a range of several GPa is seen only for the powder measurements taken without a helium pressure-transmitting medium. This is identified by the simultaneous observation of distinct diffraction peaks fit by one of the two transitioning phases. Such coexistence is not seen in the powder diffraction data measured with a helium medium, as is shown in the [supplementary information Figures ...]. Phase coexistence is not seen in the single-crystal measurements, with no observation of multiple patterns around the transition pressures.

%TODO - Check this paragraph.
A monoclinic space group for the HP-II phase has been proposed by a recent study of the compound's high-pressure properties\cite{wang2018a}. Fits to the diffraction patterns of the HP-II phase were attempted using this model, treating the observed volume collapse with a reduction of the in-plane lattice parameters. These refinements were unsatisfactory and are shown and discussed in the supplementary information. From the single-crystal data it is unambiguous that the HP-II transition involves a reduction of the $d$-spacing of the $(001)$ peak and thus the inter-planar separation. Previously proposed HP-II structures which include no such reduction are thus incompatible with the observation from these experiments. This key factor is consistent across all measurements presented here, and confirms that the correct structural model for the HP-II phase is a $P\bar{3}1m$ structure with reduced inter-planar spacing.

\subsection*{Discussion}

These results verify the occurrence of the two structural transitions under pressure in FePS$_3$. The first transition involves a shear of $\sim a/3$ along the $\mathbf{a}$ direction and the second an increase in symmetry to a $P\bar{3}1m$ space group. The results verify the model proposed by Haines et al.\cite{haines2018} attributing the volume collapse to a reduction of inter-layer separation.
%The finding of a symmetry increase and unambiguous reduction in the inter-planar spacing with the HP-I to HP-II transition is incompatible with other models proposed for this material which maintain the monoclinic space group across the entire pressure range with a discontinuous change in the $a$ and $b$ lattice parameters in HP-II\cite{wang2018a}.

The magnitudes of the lattice parameters show differences between the data sets. This is particularly evident for the powder data with a helium pressure medium in Figure \ref{fig:Fe-Fe_distance}, and is also true for data $\leq$\SI{20}{\giga\pascal} in the HP-II phase in Figure \ref{fig:SingleXtalvsPowder001}. The observed differences between the experiments suggest that both the presence of a helium pressure medium and the form of the sample used are both important factors in the high-pressure structural behaviour of FePS$_3$. There are multiple possible origins for this. One is the intended effect of the pressure-transmitting medium to allow a more hydrostatic pressure distribution in the sample. Fractions of the sample experiencing a different local pressure may explain the observation of phase coexistence only in the no medium data, but the effect of a more significant uniaxial component of pressure on the inter-planar spacing and volume collapse requires further investigation.

 A more subtle effect may be the interaction of helium itself with FePS$_3$. The intercalation of $M$PS$_3$ materials with elements such as lithium\cite{grasso2003} has been found to strongly affect both the electronic and magnetic properties. Differences in behaviour between the single crystal and powder, both with a helium medium, such as those seen in Figure \ref{fig:Fe-Fe_distance}, could then arise from differing amounts of intercalation of helium into the material due to the sample form. The impact of such factors is important for cases of other layered compounds such as the intercalation of graphite with Li\cite{guo2016} or Yb\cite{weller2005}, in which intercalation is limited to the layer edges of samples. In this way intercalation in FePS$_3$ may be more complete in the powder sample, giving rise to observed differences in behaviour.

Regardless of its microscopic origin, a change in the extent of the reduction of inter-planar spacing accompanying the symmetry increase between the experiments presented here is important. It is direct evidence that the high-pressure properties of this material and others in the same family are sensitive to experimental factors which may not be considered by calculations. These theoretical predictions themselves must make careful consideration when using previous experimental observations of crystal structure as a basis for the determination of properties such as band structure calculations.

We verify that the structural transition linked to metallisation and a new magnetic state in FePS$_3$ involves a symmetry increase from space group $C2/m$ to $P\bar{3}1m$ and that the volume collapse is due to a reduction of inter-planar spacing rather than intra-planar effects. We find that the extent of this collapse is sensitive to the presence of a helium pressure-transmitting medium, and note that consideration must be given to this such factors in future experiments and calculations.

\subsection{Acknowledgements}

This research was supported by United Kingdom Research and Innovation Global Challenges Research Fund COMPASS Grant No. ES/P010849/1 and Cambridge Central Asia Forum, Jesus College, Cambridge. This project has received funding from the UK Department of Business, Environment and Industrial Strategy to support collaboration between Cavendish Laboratory and Navoi Mining Institute. 

This work was carried out with the support of the Diamond Light Source through the award of beamtime for proposals EE15949 and CY23524.
\bibliography{FePS3SingleCrystal_new.bib}

\end{document}